\documentclass[letterpaper,11pt]{article}
\pdfoutput=1

\usepackage{jheppub}

\usepackage{multirow,subfig,xspace,xcolor}
\usepackage[countmax]{subfloat}
\usepackage{CJKutf8}
\usepackage{CJK}

\definecolor{darkblue}{rgb}{0,0,0.5}

\DeclareRobustCommand{\Sec}[1]{Sec.~\ref{#1}}

\DeclareRobustCommand{\App}[1]{App.~\ref{#1}}
\DeclareRobustCommand{\Tab}[1]{Table~\ref{#1}}

\DeclareRobustCommand{\Fig}[1]{Fig.~\ref{#1}}

\DeclareRobustCommand{\Eq}[1]{Eq.~(\ref{#1})}

\DeclareRobustCommand{\Ref}[1]{Ref.~\cite{#1}}

\title{Calculating Pull for Non-Singlet Jets}

\author{Yunjia Bao (包昀嘉)}

\author{and Andrew J.~Larkoski}
\affiliation{Physics Department, Reed College, Portland, OR 97202, USA}

\emailAdd{baoy@reed.edu}
\emailAdd{larkoski@reed.edu}

\abstract{
The pull vector is a jet observable sensitive to the distribution of soft radiation controlled by the color flow in a collider event.  We present calculations to leading order in the soft and collinear limits for the pull vector measured between pairs of jets that do not form a color-singlet dipole.  Our calculations are presented within the context of $e^+e^-\to$ three jets events, on which pull is measured between the two subleading jets.  A subset of these calculations can be re-interpreted as a bottom--anti-bottom quark jet pair in a color octet configuration, which can be a background to Higgs production at large boost.  We also present a universal expression for the pull distribution in the high-boost and small jet radius limit.  This distribution is controlled by color SU(3) quadratic Casimirs that arise from product representations of pairs of QCD jets.
}

\begin{document}
\begin{CJK*}{UTF8}{gbsn}
\maketitle
\end{CJK*}

\section{Introduction}

In the search for new physics at the Large Hadron Collider (LHC), a central goal is to measure all quantum numbers and couplings of known Standard Model particles as well as to observe as-of-yet undiscovered particles.  At a collider experiment, measuring the mass of a particle is straightforward because detectors measure nearly all of the energy produced in collision.  Techniques exist for determining a particle's electric charge by weighting hits in the tracking system by their energy \cite{Field:1977fa,Krohn:2012fg}.  However, direct measurement of a particle's charge under quantum chromodynamics (QCD) or color is subtle, as all of the particles that are actually detected by experiment are color-neutral.  A particle's color, or at least if it has non-zero color, is inferred from jet production and quantities measured on jets \cite{Hanson:1975fe,Brandelik:1979bd,Barber:1979yr,Berger:1979cj,Bartel:1979ut}.

The observable pull \cite{Gallicchio:2010sw} was introduced to be directly sensitive to the flow of color between pairs of jets.  The distribution of soft radiation throughout a collision event is determined by the location and connections of color dipoles, the ends of which are the observed jets.  Pull quantifies the location of the dominant soft radiation between two jets, thus providing a measure of their color connectedness or their color ``pull'' on one another.  Two jets that form a color-singlet dipole, from the decay of a color-singlet resonance, for example, will dominantly emit soft radiation in the region between the jets, because gluons emitted at wide angles would only see the net zero color of the two jets.  By contrast, two jets that are produced from standard QCD processes at the LHC would in general have color connections to the initial state, as well as to whatever other color objects were produced.  Thus, radiation about these jets would have a much weaker correlation with the relative locations of the jets of interest.

While pull has a very nice physical interpretation, has been studied extensively in simulation, and has even been measured in experiment \cite{Abazov:2011vh,Aad:2015lxa,Aaboud:2018ibj} and used in searches \cite{Chatrchyan:2012xdj,Chatrchyan:2013zna,Chatrchyan:2014tja,Khachatryan:2014dea,Khachatryan:2016mdm}, there has been little theoretical analysis of the observable to honestly justify that it does what it is claimed to do.  The first calculations of pull as measured on the two jets from color-singlet decay were presented in \Ref{Larkoski:2019urm}, which illustrated the challenges of the calculation and demonstrated that at least in this restricted case, pull acts as expected.  Further, pull had been measured in experiment on jets from $W$ boson decay, which enabled direct comparison to data.

Nevertheless, to demonstrate that pull is indeed sensitive to the color connection between two jets, we should demonstrate that there is a significant difference between the calculated distribution of pull for pairs of jets that do and do not form a color-singlet dipole.  This is our goal in this paper.  The simplest collision event in which pairs of jets do not form color-singlets is $e^+e^- \to$ three jet events, which is what we consider first.  For simplicity, we restrict our analysis to leading-order in the strong coupling and to leading power in the soft or collinear limits.  We will measure the color-connectedness of the two jets closest in angle with pull.  With three jets in the final state, there are therefore three distinct dipoles from which soft radiation can be emitted.  We will show that the radiation from two of these dipoles can be accounted for from calculations presented in \Ref{Larkoski:2019urm}, with some re-interpretation.  We present a new calculation for the distribution of radiation from the dipole which does not include the jet on which pull is directly measured.

Our calculations for pull in three-jet final states can then be leveraged to theoretically understand one of the original motivations for the observable.  Identifying the decay of the Higgs boson to bottom quarks at high significance is a challenge at the LHC.  Once jets have been tagged as containing bottom quarks, a major background to $H\to b\bar b$ decay is the gluon splitting process $g\to b\bar b$.  Because this splitting lacks a soft singularity, once the invariant mass of the bottom quarks is selected for, the kinematics of the bottom quarks from Higgs decay and gluon splitting are nearly identical.  However, because the Higgs boson is a color singlet, radiation from the bottom quarks is confined to lie between them, distinct from the gluon splitting case.  This suggests that pull may provide discrimination power between these two processes, though this has not been observed in simulation \cite{ATLAS:2014rpa,deOliveira:2015xxd,Lin:2018cin}.

With an explicit calculation, we are able to understand the discrimination properties of pull in a controlled, well-defined context.  Working in the limits in which both the Higgs is highly boosted and the radii of the bottom quark subjets is small, we are able to explicitly calculate the discrimination power of pull for this problem, quantified in the signal vs.~background efficiency curve.  We conjecture that in these boosted and collinear limits the pull distribution exhibits a universality, exclusively depending on the color configuration of the two nearby jets.  In these limits, we are able to write down a master formula for the distribution of pull, for any two jets in QCD on which it might be measured.  We enumerate all possible irreps of color SU(3) that arise in the product representation of the color of two QCD jets and how that affects the corresponding pull distribution.  This explicitly demonstrates that pull is indeed sensitive to the color flow between a pair of jets in a simplified limit.  We leave validation of this observation in simulation to future work.

This paper is organized as follows.  In \Sec{sec:obsdef}, we first define the pull observable, presenting a slightly modified definition from that originally proposed that is more natural in the $e^+e^-$ collision case.  In \Sec{sec:ee3jet}, we present the calculation of the distribution of the pull observable measured on the closest two jets in angle produced in $e^+e^-\to $ three jets events.  \Sec{sec:higgs} expands on these results, and applies them to the problem of discrimination of $H\to b\bar b$ and $g\to b\bar b$ in the highly boosted limit.  We explore pull for all possible representations of SU(3) color that can arise in the product of the color of two QCD jets in \Sec{sec:colorreps}, and conclude and discuss future directions in \Sec{sec:concs}. Appendices contain details of calculations quoted in the body of the paper.

\section{Observable Definitions}\label{sec:obsdef}

For application to jets produced in hadron collisions, \Ref{Gallicchio:2010sw} introduced the pull vector $\vec t$ as:
\begin{equation}\label{eq:pullorigdef}
\vec t_\text{original} = \sum_{i\in J}\frac{p_{\perp i} |\vec r_i|}{p_{\perp J}}\vec r_i\,.
\end{equation}
Here, the sum runs over the particles $i$ in a jet $J$ of interest, $p_\perp$ is the momentum transverse to the collision beam, and the vector $\vec r_i$ is
\begin{equation}
\vec r_i = (y_i-y_J,\phi_i-\phi_J)\,.
\end{equation}
The jet center is located at rapidity-azimuth of $(y_J,\phi_J)$ and particle $i$ is located at $(y_i,\phi_i)$.  The jet center is just defined as the vector sum of the momenta of all particles that compose the jet.  Pull is therefore defined as a two-dimensional vector in the plane of the cylindrical detector.  As used to probe color connections, pull can be measured on two jets and the directions of their vectors compared.  Jets with a strong color connection (i.e., jets that form a color-singlet dipole), will have pull vectors that point toward one another, while weakly color-connected jets will have pull vectors with a random relative orientation.

For the calculations presented in this paper, however, we use a slightly different definition of the pull vector motivated both by our study of jets in $e^+e^-$ collisions as well as simplifying analytical calculations.  This modified definition of the pull vector was introduced in \Ref{Larkoski:2019urm} and is
\begin{equation}
\vec t_\text{modified} = \sum_{i\in J}\frac{E_i \sin^2\theta_i}{E_J}(\cos\phi_i,\sin\phi_i)\,.
\end{equation}
Now, $E$ is the energy, $\theta_i$ is the angle from particle $i$ to the jet center, and $\phi_i$ is the azimuthal angle of the particle about the jet center.  In the collinear limit, these two definitions are identical, but in general differ at finite angle. As our investigation mainly concerns with the limit where the jet radius $R \ll 1$, this modified definition of the pull vector simplifies the calculation without losing the limit behavior of interest. For all results presented in this paper, the azimuthal angle $\phi_i$ will be defined with respect to the location of a neighboring jet's center.  Specifically, if a particle $i$ lies on the line between the jet $J$ and the reference jet, $\phi_i = 0$, while if it is on the other side of jet $J$, $\phi_i = \pi$.  This is what we will mean by measuring pull on a pair of jets: we calculate the pull vector of one jet whose components are defined with respect to the location of the second jet.

Rather than the Cartesian components of the pull vector, we will typically express it as its magnitude $t$ and azimuthal angle $\phi_p$.  We call $\phi_p$ the pull angle and it is defined as
\begin{equation}
\phi_p = \cos^{-1}\frac{t_x}{t} = \cos^{-1}\frac{\sum_{i\in J}E_i \sin^2\theta_i\, \cos\phi_i}{\left|
\sum_{i\in J}E_i \sin^2\theta_i\,(\cos\phi_i,\sin\phi_i)
\right|}\,.
\end{equation}
The pull vector is infrared and collinear (IRC) safe, and so its distribution can be calculated order-by-order in perturbation theory.  However, the pull angle $\phi_p$ alone is not IRC safe.  We will have to deal with this later as the pull angle is the aspect of the pull vector that is most sensitive to color connections between jets.

\section{Pull in $e^+e^-\to 3$ jet events}\label{sec:ee3jet}

\begin{figure}[t]
\begin{center}
\includegraphics[width=7cm]{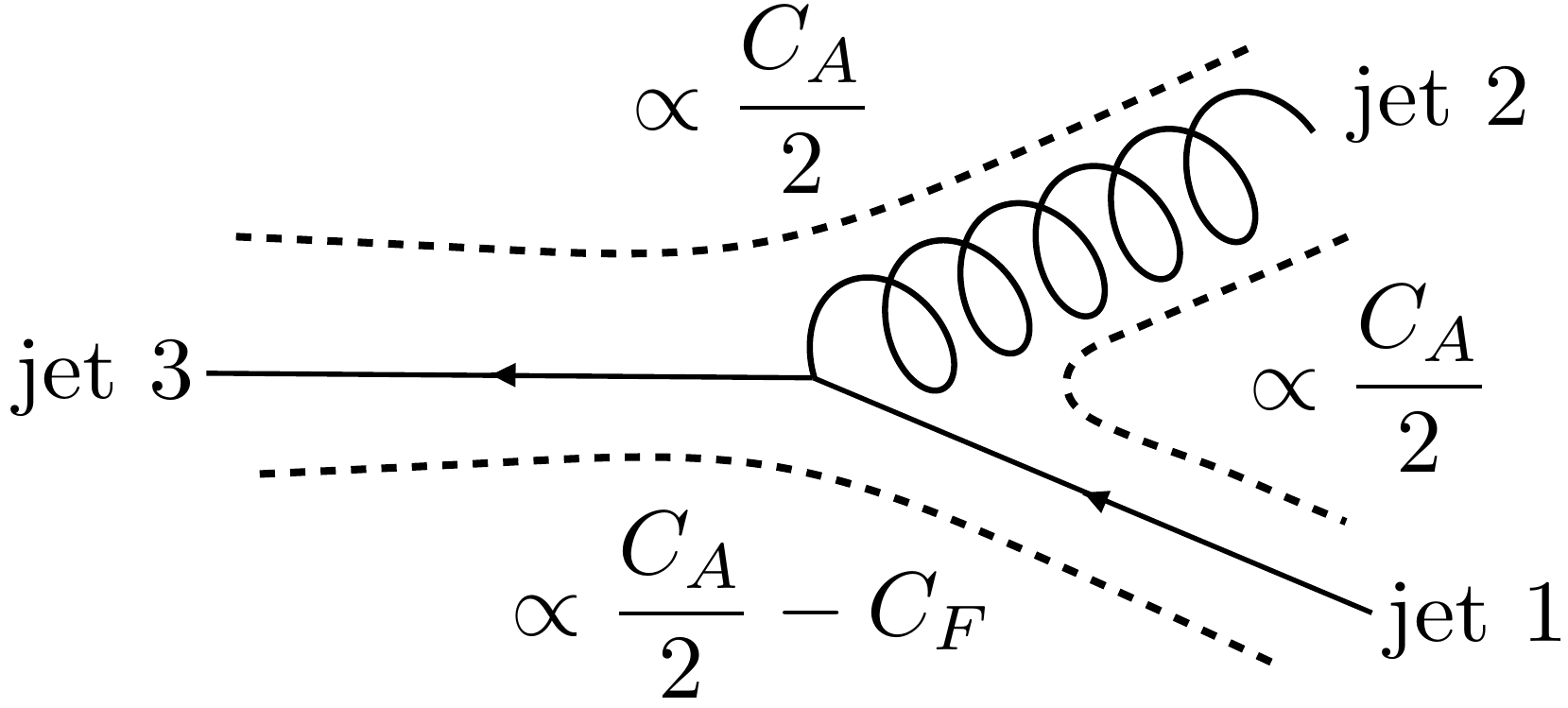}
\end{center}
\caption{Illustration of color flow between the three jets produced in $e^+e^-\to q\bar q g$ collisions.  The dashed lines represent the flow of color in dipoles that stretch between pairs of jets with the corresponding absolute value of the color charge carried by each dipole indicated.
}
\label{fig:qgq_dipoles}
\end{figure}

To illustrate the form of the pull distribution for pairs of jets that do not form a color-singlet dipole, we will study pull as measured on pairs of jets in $e^+e^-\to$ three jets events.  The setup of how we measure pull on this final state is illustrated in \Fig{fig:qgq_dipoles}.  For a three-jet final state in the center-of-mass frame, those three jets lie in a plane and the most energetic will be isolated in a hemisphere about the collision point.  We refer to this most energetic jet as jet 3.  The two lower-energy jets are the pair closest in angle.  Of these two jets, jet 1 is the most energetic and the jet on which we measure pull.  Jet 2, the lowest-energy jet, defines the axis along which the pull angle is defined to be 0.  \Fig{fig:qgq_dipoles} shows that the quark is the most energetic jet, and the anti-quark is the second most-energetic jet.  Thus, in this configuration, we measure pull on the anti-quark jet with respect to the gluon jet's direction.

\Fig{fig:qgq_dipoles} also shows the strength of color correlation between the pairs of jets, as measured by the product of each particle's color matrix. Because the total color of the final state is 0, the sum of the color of the quark, anti-quark, and gluon is 0:
\begin{equation}\label{eq:color0}
{\bf T}_q+{\bf T}_{\bar q}+{\bf T}_g=0\,.
\end{equation}
The square of any of the color matrices is just the quadratic Casimir for that particular color representation.  In QCD, we have
\begin{align}
&{\bf T}_q^2={\bf T}_{\bar q}^2 = C_F = \frac{4}{3}\,, &{\bf T}_g^2 = C_A = 3\,.
\end{align}
By dotting the individual color matrices with \Eq{eq:color0}, we can solve for the values of the dot products of pairs of color matrices, that correspondingly determine the strength of color connectedness of two jets.  We have
\begin{align}
&{\bf T}_q\cdot {\bf T}_{\bar q} = \frac{C_A}{2}-C_F\,, &{\bf T}_q\cdot {\bf T}_g= {\bf T}_{\bar q}\cdot {\bf T}_g=-\frac{C_A}{2}\,.
\end{align}

With this setup, we would like to calculate the differential cross section of the pull vector, or equivalently, the double differential cross section of the pull magnitude $t$ and the pull angle $\phi_p$.  To do this calculation, we will work to lowest order in the strong coupling $\alpha_s$ and in the leading soft or collinear limits, where the pull magnitude is small, $t\ll 1$.  With these approximations, the double differential cross section decomposes into a sum of contributions from soft and collinear emissions:
\begin{equation}
\frac{d^2\sigma^{t\ll 1}}{dt\, d\phi_p}=S_{q\bar q g}(t,\phi_p) + J(t,\phi_p)\,.
\end{equation}
Here, we refer to $S_{q\bar q g}(t,\phi_p)$ as the soft function for pull and $J(t,\phi_p)$ as the jet function for pull which encode the contribution to pull from soft and collinear emissions, respectively.  Note that no rigorous, all-orders factorization of the pull cross section is assumed or implied here; this decomposition simply follows from the factorization of QCD matrix elements into these components.

The separation of the emission phase space into soft and collinear regions is arbitrary, so one needs to use some regularization scheme to do it.  For pull, dimensional regularization is sufficient to uniquely define the soft and jet functions individually.  Thus, calculation of the soft function, for example, proceeds by calculating the distribution of pull on the dimensionally-regulated phase space with the single-emission, eikonal matrix element.  There are three possible dipoles off of which a soft particle can be emitted, and we need to sum together each of their contributions.  We won't present explicit calculations in the text here, but will discuss how the pieces fit together for the complete soft function.

Following the identification of jets in \Fig{fig:qgq_dipoles}, one of the dipoles that can emit a soft gluon is the dipole formed from jets 1 and 2.  The calculation of this contribution to the soft function was done in \Ref{Larkoski:2019urm}, with the only necessary change to the case at hand to replace the Casimir $C_F$ in that calculation with $(-{\bf T}_1\cdot {\bf T}_2)$ for the three-jet case.  Next, the soft gluon could be emitted off of the dipole formed by jets 1 and 3.  Note that, relative to jet 2, jet 3 has an azimuthal angle $\pi$ about jet 1, by momentum conservation.  Also, this dipole still contains the jet 1, on which pull is measured, and so we can use the results of \Ref{Larkoski:2019urm} again, with two modifications.  First, the Casimir $C_F$ in the soft function result of \Ref{Larkoski:2019urm} should be replaced by $(-{\bf T}_1\cdot {\bf T}_3)$ for the three-jet case.  Second, the pull angle $\phi_p$ should be rotated by $\pi$ to represent the orientation of jet 3 with respect to jet 2.  The contribution of soft gluons emitted off of the dipole formed from jets 2 and 3 is novel, and requires a new calculation.  For this case, the dipole that emits the soft gluon does not contain the jet on which pull is measured.  As such, this contribution to the soft function lacks a collinear singularity.  This calculation is presented in \App{app:nonconsoft}.

Adding together all three possible sources of soft radiation, the leading-order soft function for pull when $t>0$ measured on $e^+e^-\to$ three jet events is:
\begin{align}\label{eq:softfunctot}
S_{q\bar q g}(t,\phi_p) &= \frac{\alpha_s}{\pi^2}\frac{1}{t}\left[
{\bf T}_1^2\log\frac{\mu^2 \tan^2\frac{R}{2}}{t^2 E_J^2 \sin^2\phi_p}\right.\\
&
\hspace{1cm}
+(-{\bf T}_1\cdot {\bf T}_2)f(\phi_p,\theta_{12})+(-{\bf T}_1\cdot {\bf T}_3)f(\pi+\phi_p,\theta_{13})+(-{\bf T}_2\cdot {\bf T}_3)g(\phi_p)
\biggr]\,.\nonumber
\end{align}
Here, $\mu$ is the dimensional regularization scale and the function $f(\phi_p,\theta)$ was calculated in \Ref{Larkoski:2019urm} and is
\begin{align}
f(\phi_p,\theta)=2\cot \phi_p\,\, \tan^{-1}\frac{\frac{\tan\frac{R}{2}}{\tan\frac{\theta}{2}}\sin\phi_p}{1-\frac{\tan\frac{R}{2}}{\tan\frac{\theta}{2}}\cos\phi_p}-\log\left(
1+\frac{\tan^2\frac{R}{2}}{\tan^2\frac{\theta}{2}}-2\frac{\tan\frac{R}{2}}{\tan\frac{\theta}{2}}\cos\phi_p
\right) \,.
\end{align}
The function $g(\phi_p)$ is calculated in \App{app:nonconsoft} and is
\begin{align}
g(\phi_p) &= \frac{(\tan\frac{\theta_{12}}{2} + \tan\frac{\theta_{13}}{2})^2}{\tan^2\frac{\theta_{12}}{2}+\tan^2\frac{\theta_{13}}{2}+2\tan\frac{\theta_{12}}{2}\tan\frac{\theta_{13}}{2}\cos(2\phi_p)}\\
&
\hspace{1cm}
\times\left[
\frac{\sin\frac{\theta_{12}-\theta_{13}}{2}}{\sin\frac{\theta_{12}+\theta_{13}}{2}}\log\left(
\frac{\frac{\tan^2\frac{R}{2}}{\tan^2\frac{\theta_{13}}{2}}+1+2\frac{\tan\frac{R}{2}}{\tan\frac{\theta_{13}}{2}}\cos\phi_p}{\frac{\tan^2\frac{R}{2}}{\tan^2\frac{\theta_{12}}{2}}+1-2\frac{\tan\frac{R}{2}}{\tan\frac{\theta_{12}}{2}}\cos\phi_p}
\right)\right.\nonumber\\
&\left.
\hspace{2cm}
+2\cot\phi_p \tan^{-1}\left(
\frac{\left(
\frac{\tan\frac{R}{2}}{\tan\frac{\theta_{12}}{2}}-\frac{\tan\frac{R}{2}}{\tan\frac{\theta_{13}}{2}}+2\frac{\tan\frac{R}{2}}{\tan\frac{\theta_{12}}{2}}\frac{\tan\frac{R}{2}}{\tan\frac{\theta_{13}}{2}} \cos\phi_p
\right)\sin\phi_p}{1-\left(
\frac{\tan\frac{R}{2}}{\tan\frac{\theta_{12}}{2}}-\frac{\tan\frac{R}{2}}{\tan\frac{\theta_{13}}{2}}
\right)\cos\phi_p-\frac{\tan\frac{R}{2}}{\tan\frac{\theta_{12}}{2}}\frac{\tan\frac{R}{2}}{\tan\frac{\theta_{13}}{2}}\cos(2\phi_p)}
\right)
\right]\,.\nonumber
\end{align}
$\theta_{ij}$ is the angle between jets $i$ and $j$. For the coefficient of the logarithmic term in \Eq{eq:softfunctot} above, we have used the conservation of color to express
\begin{equation}
-{\bf T}_1\cdot {\bf T}_2+
-{\bf T}_1\cdot {\bf T}_3 = {\bf T}_1\cdot \left(
-{\bf T}_2-{\bf T}_3
\right) = {\bf T}_1^2\,.
\end{equation}

For a general configuration of the three final state jets, jet 1, on which pull is measured, can be any of the quark, anti-quark, or gluon jet.  For the collinear emission contribution to the cross section, we then need to calculate pull as measured on either collinear emissions from quark jets or from gluon jets.  The calculation of the quark jet function for pull was presented in \Ref{Larkoski:2019urm}, while the gluon jet function is novel, and its calculation is presented in \App{app:gluejet}.  We can express the jet function for either quarks or gluons as
\begin{equation}
J(t,\phi_p)=\frac{\alpha_s {\bf T}_1^2}{\pi^2}\frac{1}{t}\left[
\log\frac{4t E_J^2 \sin^2\phi_p}{\mu^2}-B_1
\right]\,,
\end{equation}
where $B_1$ comes from hard collinear splittings and is
\begin{align}
&B_q = \frac{3}{4}\,, &B_g = \frac{11}{12} - \frac{n_f }{6C_A}\,,
\end{align}
for quark and gluon jets, respectively.  Other than the (trivial) $\sin^2\phi_p$ dependence in the logarithm, collinear emissions are flat in $\phi_p$: they are at too small of an angle to know the specific direction of jet 2 and are uncorrelated with any other jets in the event.

Adding the soft and jet functions together, we find the cross section for pull at leading order to be
\begin{align}\label{eq:gen3jet}
\frac{d^2\sigma^{t\ll 1}}{dt\, d\phi_p} &=\frac{\alpha_s }{\pi^2}\frac{1}{t}\left[
{\bf T}_1^2\log\frac{4 \tan^2\frac{R}{2}}{t}-{\bf T}_1^2 B_1\right.\\
&
\hspace{2cm}
+(-{\bf T}_1\cdot {\bf T}_2)f(\phi_p,\theta_{12})+(-{\bf T}_1\cdot {\bf T}_3)f(\pi+\phi_p,\theta_{13})+(-{\bf T}_2\cdot {\bf T}_3)g(\phi_p)
\biggr]\,.\nonumber
\end{align}
Note that the dimensional regularization scale $\mu$ has dropped out; the physical cross section is independent of this unphysical scale.

\subsection{Inclusive Prediction for Pull}
With the differential cross section of the pull vector, we can assume that the soft and collinear contribution of the pull vector factorizes from an $e^+e^- \to q\bar{q}g$ event. Then, the soft and jet functions are regarded as a conditional probability density of the pull vector for observing an additional soft or collinear emission off from one of the three particles in the final state. Thus, the inclusive differential cross section of the pull vector for an $e^+e^- \to q\bar{q}g+X$ event can be found by integrating over the phase space of the final state particles
\begin{equation}
\frac{d^2\sigma_{q\bar{q}g}}{dt\, d\phi_p} = \int_0^1 d x_q\, \int_0^1 d x_{\bar{q}}\, \Theta(x_q + x_{\bar{q}} - 1)\,\frac{1}{\sigma_0} \frac{d^2\sigma(e^+e^- \to q\bar{q}g)}{d x_q\, d x_{\bar{q}}} \sum_\text{jet orderings} \frac{d^2\sigma^{t\ll 1}}{dt\, d\phi_p}\, \Theta_\text{pull}\,,
\label{eq:3jetincl}
\end{equation}
in which $\frac{d^2 \sigma(e^+e^- \to q\bar{q}g)}{d x_q\, d x_{\bar{q}}}$ denotes the cross section for $e^+e^- \to q\bar{q}g$ in terms of energy fractions $x_q$ and $x_{\bar{q}}$, and $\Theta_\text{pull}$ is the phase space constraints for identifying the two jets on which pull is measured. The differential cross section for $e^+e^- \to q\bar{q}g$ is 
\begin{equation}
\frac{1}{\sigma_0}\frac{d^2 \sigma(e^+e^- \to q\bar{q}g)}{d x_q\, d x_{\bar{q}}} = \frac{\alpha_s C_F}{2\pi} \frac{x_q^2 + x_{\bar{q}}^2}{(1 - x_q)(1 - x_{\bar{q}})}\,,
\end{equation}
in which the energy fraction $x_i$ of particle $i$ is defined as 
\begin{align}
&x_i = \frac{2p_i \cdot Q}{Q^2} \,,  &Q = p_1 + p_2 + p_3\,.
\end{align}
%Note that the three energy fractions sum up to $2$ instead of $1$. 
Note that the angular dependence in the jet and soft function frequently appears as $\tan(\theta_{ij}/2)$. A straightforward algebraic manipulation with dot products shows that 
\begin{equation}
	\tan\left(\frac{\theta_{ij}}{2}\right) = \sqrt{\frac{1 - \cos\theta_{ij}}{1 + \cos\theta_{ij}}} = \sqrt{\frac{x_i + x_j - 1}{(1-x_i )(1-x_j )}}.
\end{equation}
where $\theta_{ij} \in [0, \pi]$. 

With $\Theta_\text{pull}$, we aim to describe the following algorithms for the pull measurement: (1) identify the two lower energy jets, (2) identify the one of them with a higher energy, and (3) ensure that those two jets are separated by at least $2R$ such that they are recognized as distinct jets. Therefore, we demand that the following three inequalities hold:
\begin{equation}
E_1 > E_2, \qquad E_3 > E_1, \qquad \theta_{12} > 2R. 
\end{equation}
These can correspondingly be expressed as in terms of the three-body final state energy fractions $x_i$:
\begin{equation}
\Theta_\text{pull} =\Theta(x_3 - x_1) \Theta(x_1 - x_2)\Theta\left(\frac{x_1 + x_2 - 1}{(1-x_1)(1-x_2)} - \tan^2 R\right) \Theta(x_2 - x_\text{cut})\,.
\end{equation}
The rightmost $\Theta$-function enforces $x_2>x_\text{cut}$ and ensures that the cross section is IRC safe.

\begin{figure}[t!]
	\begin{center}
		\raisebox{0.3cm}{\includegraphics[height=6.4cm]{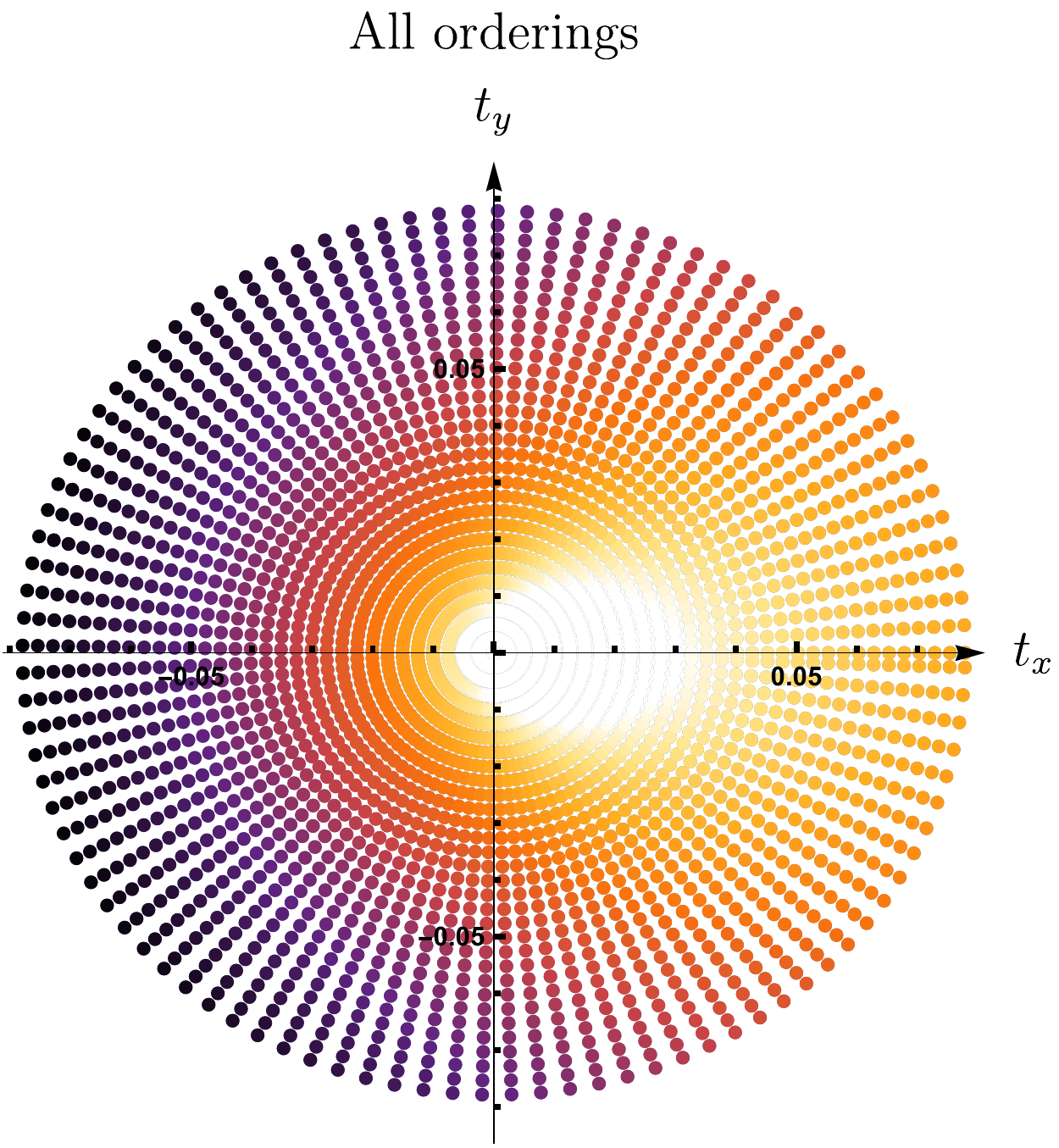}}
		\hspace{0.75cm}
		\includegraphics[height=7cm]{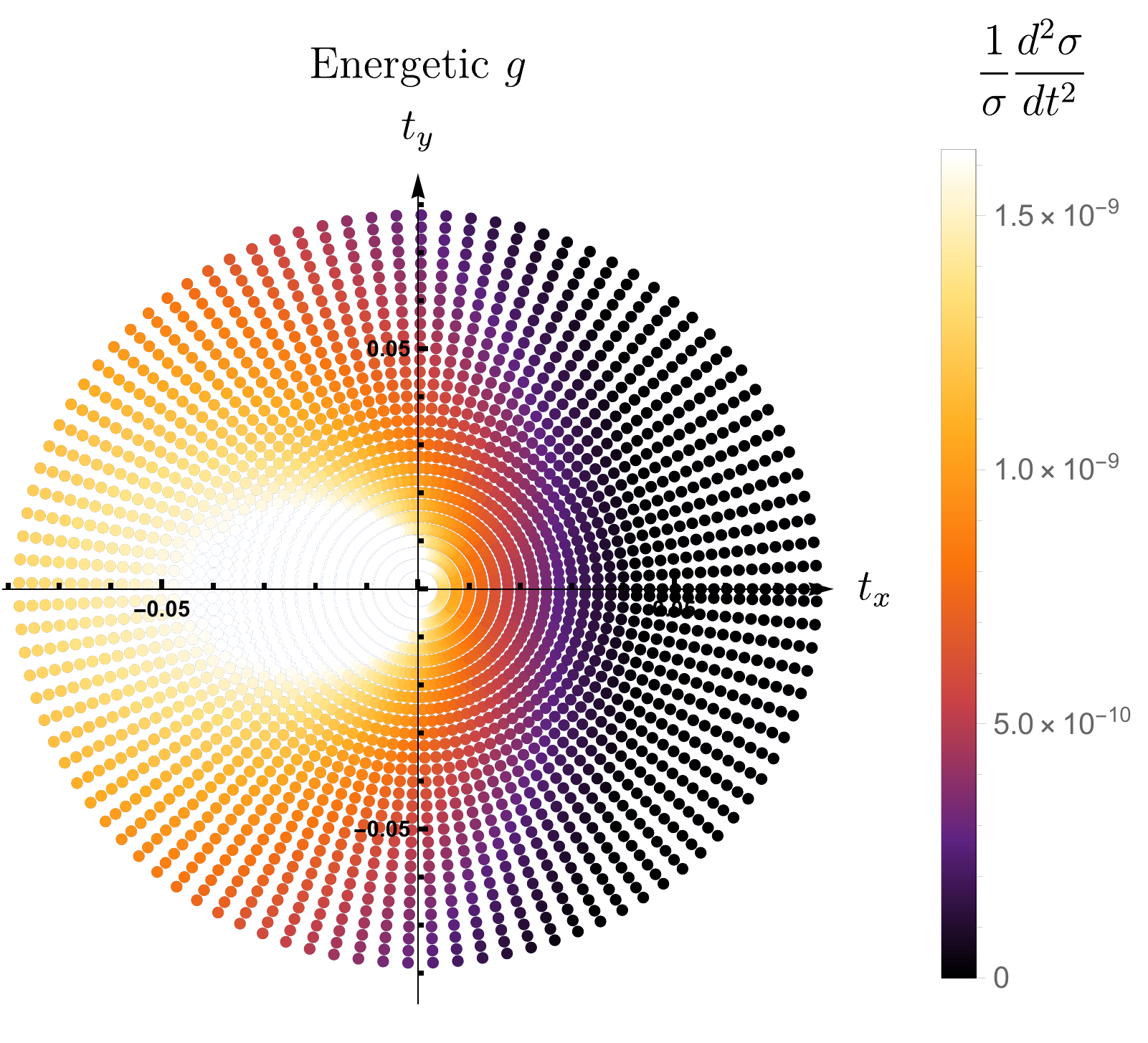}
	\end{center}
	\caption{Plots of the relative double differential cross section of the pull vector measured on $e^+e^-\to q\bar q g$ events in the $(t_x, t_y) = (t \cos\phi_p, t \sin\phi_p)$ plane.  The colors range over a linear scale with white (black) corresponding to the largest (smallest) cross section. To make these plots, we have set the jet radius $R = 0.5$, the jet energy cut $x_\text{cut} = 0.2$ and the number of active fermions $n_f = 5$.  (left) Pull vector summed over all six jet orderings. (right) Pull vector when gluon is the most energetic jet. 
	}
	\label{fig:double_cross_section}
\end{figure}

With these phase space restrictions and matrix elements identified, we can then perform the integrals in \Eq{eq:3jetincl}, summing over all possible orderings of the $q$, $\bar q$, and $g$ jets.  The result of this calculation is shown on the left in \Fig{fig:double_cross_section}.  In that figure, we have plotted the relative differential cross section, in terms of the components of the pull vector
\begin{align}
&t_x = t\cos\phi_p\,, & t_y = t\sin\phi_p\,.
\end{align}
Recall that $\phi_p = 0$ is in the direction of the nearby reference jet.  To make this plot, we have set the jet radius $R = 0.5$, the jet energy cut value $x_\text{cut} = 0.2$, and the number of active fermions to $n_f = 5$.  The cross section is highly peaked about the positive $t_x$ axis, demonstrating that most radiation is present in the region between the two jets.  This is to be expected: because of the soft and collinear singularities of the gluon in the $e^+e^-\to q\bar q g$ matrix element, the dominant configuration of the jets is with the gluon as the lowest energy jet.  Therefore, pull is measured about the quark or anti-quark jet, with respect to the direction of the gluon jet, and this pair of jets lives in the {\bf 3} or $\bar {\bf 3}$ representation of SU(3) color.  The product of color factors in this configuration is $-{\bf T}_1\cdot {\bf T}_2 = C_A/2>0$, and so the cross section peaks around $t_y = 0$ and $t_x> 0$ ($\phi_p = 0$) because the functions $f(\phi_p,\theta)$ and $g(\phi_p)$ have maxima at $\phi_p = 0$.

It is also interesting to restrict to studying the non-dominant jet configuration, by forcing the gluon jet to have the largest energy of all three final state jets.  In this configuration, the jets on which pull is measured, the quark and anti-quark, are in the {\bf 8} representation of SU(3) color.  As such, the product of their color matrices is negative, $-{\bf T}_q\cdot {\bf T}_{\bar q} = C_F - C_A/2 = -1/6$, and so the radiation is dominantly outside of the two jets on which pull is measured. This is illustrated in the plot on the right of \Fig{fig:double_cross_section}.  The cross section in this configuration is peaked about the negative $t_x$ axis, as expected.

\section{$H\to b\bar b$ vs.~$g\to b\bar b$}\label{sec:higgs}

The results presented in the previous section, along with prior calculations \cite{Larkoski:2019urm}, can inform the use of pull for identification of $H\to b\bar b$ decays.  One of the original motivations for pull presented in \Ref{Gallicchio:2010sw} was that it could be used to identify Higgs decays to bottom quarks from the dominant background of gluon splitting to bottom quarks.  Because there is no soft singularity for $g\to b\bar b$ splitting, the kinematics of $H\to b\bar b$ and $g\to b\bar b$ are nearly identical, once the mass of the pair of bottom quarks is fixed.  Thus, observables sensitive to jet kinematics, such as subjet energy fractions, are not useful for this problem.  However, the Higgs boson is a color singlet, while the gluon is a color octet, and this distinction is imprinted on the distribution of soft radiation within and about the pair of bottom quarks.  Pull is explicitly sensitive to the orientation of soft radiation, and so can be used to improve identification of Higgs decays.

In this section, we will study the discrimination power of the pull angle $\phi_p$ for Higgs decays to bottom quarks.  We will work in the highly-boosted limit in which the energy or transverse momentum of the Higgs boson is much larger than its mass $m_H$, so that the bottom quark jets are relatively collimated.  Further, we assume that the radii of the individual bottom quark jets $R$ is significantly smaller than the angular separation of the bottom quarks.  These limits are relevant and can easily be borne out in practice.  For example, for a Higgs boson with $p_\perp = 250$ GeV, the angular separation of the bottom quarks $\theta_{12}$ is approximately
\begin{equation}
\theta_{12} \simeq \frac{2m_H}{p_\perp} \simeq 1\,.
\end{equation}
Correspondingly, subjet radii for individual bottom quarks of $R\lesssim 0.4$ are reasonable as now even down to $R = 0.2$ is used in experiment \cite{Aad:2019uoz}.  Importantly, for the subjets to be well-defined and non-overlapping, their jet radius $R$ should be less than half of the angular separation of the bottom quarks, $\theta_{12}$.  Then, taking the $R\ll \theta_{12}\ll 1$ limits of the expression for the pull distribution for a color singlet from \Ref{Larkoski:2019urm} we find
\begin{align}
\frac{d^2\sigma^{H\to b \bar b,R\ll\theta_{12}\ll1}}{dt\, d\phi_p} &=\frac{\alpha_s C_F}{\pi^2}\frac{1}{t}\left[\log\frac{R^2}{t}-\frac{3}{4}
+4\frac{R}{\theta_{12}}\cos\phi_p \right]+{\cal O}\left(R^2\right)\,,
\end{align}
where $\phi_p\in[0,\pi]$.

\begin{figure}
\begin{center}
\includegraphics[width=7cm]{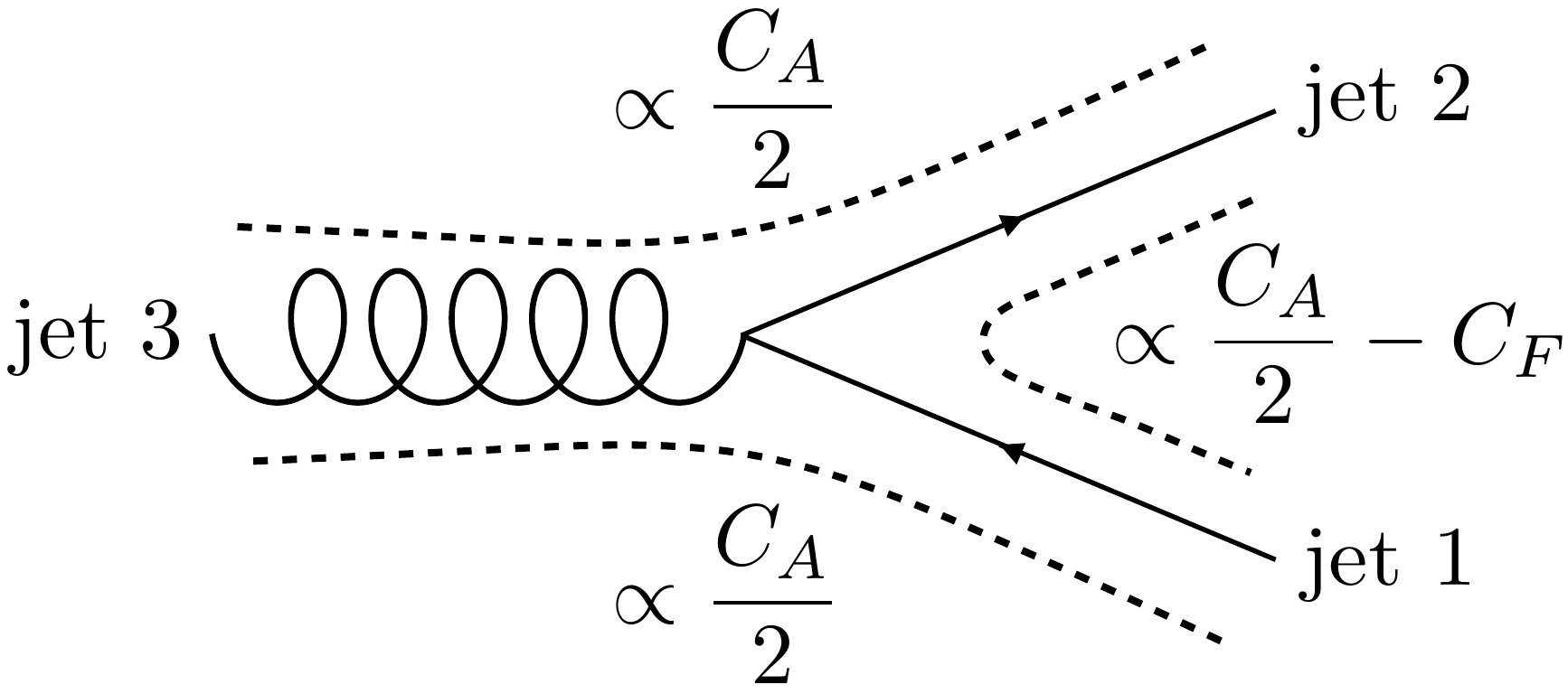}
\end{center}
\caption{Color flow in the $e^+e^-\to$ three-jet configuration relevant for comparison to boosted $H\to b\bar b$ in which the two quarks form a color octet.  Pull is measured about jet 1 with respect to jet 2.
}
\label{fig:gqq_dipoles}
\end{figure}

For the color octet configuration, we use the more general expression presented in \Eq{eq:gen3jet}.  The configuration of jets we consider produced from $e^+e^-$ collisions is illustrated in \Fig{fig:gqq_dipoles} where we restrict to the configuration in which the quark and anti-quark jets are closest in angle.  For expanding to linear order in the jet radius $R$, we first note that the expansion of the function $g(\phi_p)$ is proportional to $R^2$.  This contribution comes from a dipole that does not include the jet of interest, so for radiation from this dipole to land in the jet, it must hit an uncorrelated region of area $R^2$.  Continuing, the contribution proportional to ${\bf T}_1\cdot {\bf T}_3$ expands to linear order in $R$ as
\begin{equation}
\left.f(\pi+\phi_p,\theta_{13})\right|_{R\ll \theta_{13}} = 2R\cot\frac{\theta_{13}}{2}\cos(\pi+\phi_p) +{\cal O}\left(R^2\right)= -2R\cot\frac{\theta_{13}}{2}\cos(\phi_p) +{\cal O}\left(R^2\right)\,.
\end{equation}
In the high-boost limit of the two bottom quark jets, $\theta_{13}\to \pi$, and so $\cot\frac{\theta_{13}}{2}\to 0$.  Therefore, this term is also ignorable to linear order in the jet radius $R$.  The only relevant term in the cross section in these limits is proportional to ${\bf T}_1\cdot {\bf T}_2$, which is just the product of the color matrices of the bottom quarks.  The relevant color factor is 
\begin{align}
&{\bf T}_q\cdot {\bf T}_{\bar q} = \frac{C_A}{2}-C_F\,.
\end{align}
Therefore, the pull distribution for the color octet configuration in the small $R$ limit is
\begin{align}
\frac{d^2\sigma^{g\to b\bar b,R\ll\theta_{12}\ll1}}{dt\, d\phi_p} &=\frac{\alpha_s C_F}{\pi^2}\frac{1}{t}\left[\log\frac{R^2}{t}-\frac{3}{4}
-4\left(\frac{C_A}{2C_F}-1\right)\frac{R}{\theta_{12}}\cos\phi_p \right]+{\cal O}\left(R^2\right)\,.
\end{align}
This has been written with an explicit negative sign in front of the linear in $R$ term because $C_A/(2C_F)-1 = 1/8 > 0$.  Going forward, we will drop the remainder ${\cal O}(R^2)$ as it will be implicit in the following.

\subsection{Discrimination Power of Pull}

These distributions are already informative, but we would like to determine the distribution of the pull angle $\phi_p$ alone to identify its power as a discrimination observable.  Because the pull angle is not IRC safe, we cannot determine the distribution of $\phi_p$ by just integrating these distributions over $t$.  However, $\phi_p$ is Sudakov safe \cite{Larkoski:2013paa,Larkoski:2015lea,Larkoski:2019urm}, and so we can calculate its distribution by marginalizing against the probability distribution of the pull magnitude, $t$.  That is, for probability distribution $p(t)$ and conditional probability distribution $p(\phi_p|t)$, the distribution of the pull angle is
\begin{equation}\label{eq:ircunsafe}
p(\phi_p) = \int dt\, p(t)\, p(\phi_p|t)\,.
\end{equation}
As long as $p(t)$ has no support around $t = 0$, this integral is finite.  This is indeed the case for the physical (or resummed) distribution, so \Eq{eq:ircunsafe} provides a way to define the distribution of $\phi_p$.  Here, we will just calculate the conditional probability distribution $p(\phi_p|t)$ to lowest order in the limits we have discussed.

To lowest order, the conditional distribution $p(\phi_p|t)$ is just the ratio of the double differential cross section of $t$ and $\phi_p$ to the cross section for $t$ alone:
\begin{equation}
p(\phi_p|t) = \frac{\frac{d^2\sigma}{dt\, d\phi_p}}{\frac{d\sigma}{dt}}\,.
\end{equation}
Above, we had calculated the distributions for both the $H\to b\bar b$ and $g\to b\bar b$ configurations.  To determine the distribution for $t$ exclusively, we can just integrate over $\phi_p$.  For either the singlet or octet color configurations, the result is the same in the limits in which we work:
\begin{equation}
\frac{d\sigma}{dt} = \int_0^\pi d\phi_p \, \frac{d^2\sigma}{dt\, d\phi_p} = \frac{\alpha_s C_F}{\pi}\frac{1}{t}\left[
\log\frac{R^2}{t}-\frac{3}{4}
\right]\,.
\end{equation}
Note crucially that the term linear in $R$ integrates to 0.  From the expressions for the double differential distributions above, the conditional distributions for the Higgs decay and gluon splitting are:
\begin{align}
p_{H\to b\bar b}(\phi_p|t) &=\frac{\frac{d^2\sigma^{H\to b\bar b}}{dt\, d\phi_p}}{\frac{d\sigma}{dt}}=\frac{1}{\pi}+\frac{4}{\pi}\frac{1}{\log\frac{R^2}{t}-\frac{3}{4}}\frac{R}{\theta_{12}}\cos\phi_p\,,\\
p_{g\to b\bar b}(\phi_p|t) &=\frac{\frac{d^2\sigma^{g\to b\bar b}}{dt\, d\phi_p}}{\frac{d\sigma}{dt}}=\frac{1}{\pi}-\frac{4}{\pi}\frac{\frac{C_A}{2C_F}-1}{\log\frac{R^2}{t}-\frac{3}{4}}\frac{R}{\theta_{12}}\cos\phi_p\,.
\end{align}

We can then determine the distribution of the pull angle.  For the Higgs decay, for example, we have
\begin{align}
p_{H\to b\bar b}(\phi_p) = \int dt\, p_{H\to b\bar b}(t)\, p_{H\to b\bar b}(\phi_p|t)\,.
\end{align}
We are restricting our analysis to linear order in the jet radius $R$, which will dramatically simplify what follows.  As shown above, the pull magnitude distribution $p(t)$ actually has no contribution to it that is linear in $R$. Thus, in this integral, we only need to keep the terms in $p(t)$ at leading order in the $R\to 0$ limit.  This is correspondingly the collinear limit in which the only thing that the pull magnitude depends on is the flavor of the jet of interest.  For both $H\to b\bar b$ and $g\to b\bar b$, the jet of interest is always a quark, and so the distribution of $t$ is identical for the two processes, up to corrections of order $R^2$:
\begin{equation}
p_{H\to b\bar b}(t) = p_{g\to b\bar b}(t) +{\cal O}(R^2)\,.
\end{equation}
To the order we work, we can then safely set $p_{H\to b\bar b}(t) = p_{g\to b\bar b}(t) \equiv p(t)$, independent of production process.

With this simplification, it follows that the pull angle distribution for Higgs decay is
\begin{align}
p_{H\to b\bar b}(\phi_p) &= \int dt\, p(t)\, p_{H\to b\bar b}(\phi_p|t)= \int dt\, p(t)\, \left(
\frac{1}{\pi}+\frac{4}{\pi}\frac{1}{\log\frac{R^2}{t}-\frac{3}{4}}\frac{R}{\theta_{12}}\cos\phi_p
\right)\\
&=\frac{1}{\pi}+d_0 \cos\phi_p
\nonumber\,.
\end{align}
Note that the distribution $p(t)$ is normalized and integrates to 1, by definition.  We define $d_0$ as the corresponding moment of the pull magnitude distribution:
\begin{equation}
d_0=\frac{4}{\pi}\frac{R}{\theta_{12}}\int dt \, \frac{p(t)}{\log\frac{R^2}{t}-\frac{3}{4}}\,.
\end{equation}
With this notation, it then follows that the gluon splitting distribution is
\begin{equation}
p_{g\to b\bar b}(\phi_p)= \frac{1}{\pi}-d_0\left(\frac{C_A}{2C_F}-1\right)\cos\phi_p\,.
\end{equation}

We can estimate the value of $d_0$ by determining the mean value of the pull magnitude, $\langle t\rangle$.  In the collinear limit for a quark jet, this is at lowest order
\begin{align}
\langle t\rangle &= \frac{\alpha_s C_F}{2\pi}\int_0^1dz\, \int_0^{R^2} \frac{d\theta^2}{\theta^2} \frac{1+(1-z)^2}{z}\, z(1-z)|1-2z|\theta^2\\
&=\frac{13}{64}\frac{\alpha_sC_F}{\pi} R^2
\,.\nonumber
\end{align}
Further, the angle between the two quark jets $\theta_{12}$ can be approximated from the mass and energy of the singlet resonance.  In the collinear or high-boost limit, we have
\begin{equation}
\theta_{12} \simeq \frac{2m_H}{p_\perp}\,,
\end{equation}
where $m_H$ is the mass of the Higgs and $p_\perp$ is its transverse momentum, assuming it is central in the detector. 

Using these results and assuming that the distribution $p(t)$ is highly peaked around its mean, we then have that
\begin{align}
d_0 &= \frac{4}{\pi}\frac{R}{\theta_{12}}\int dt\, p(t) \, \frac{1}{\log\frac{R^2}{t}-\frac{3}{4}} \simeq \frac{4}{\pi}\frac{1}{\log\frac{R^2}{\langle t\rangle}-\frac{3}{4}}\frac{R}{\theta_{12}}\\
&=\frac{4}{\pi}\frac{1}{\log\left(\frac{64}{13}\frac{\pi}{\alpha_s C_F}\right)-\frac{3}{4}}\frac{Rp_\perp}{2m}\,.\nonumber
\end{align}
Evaluating everything except for the jet radius, transverse momentum, and mass, this is approximately
\begin{equation}
d_0\simeq 0.2\frac{Rp_\perp}{m_H}\,,
\end{equation}
where we have used $\alpha_s = 0.1$.  Recall that for the two bottom quark jets to be non-overlapping, we require that $R \lesssim m_H/p_\perp$, less than approximately half of the angle between the bottom quark jets.  So, $d_0$ is bounded from above by about $0.2$.

\begin{figure}[t!]
\begin{center}
\includegraphics[width=7cm]{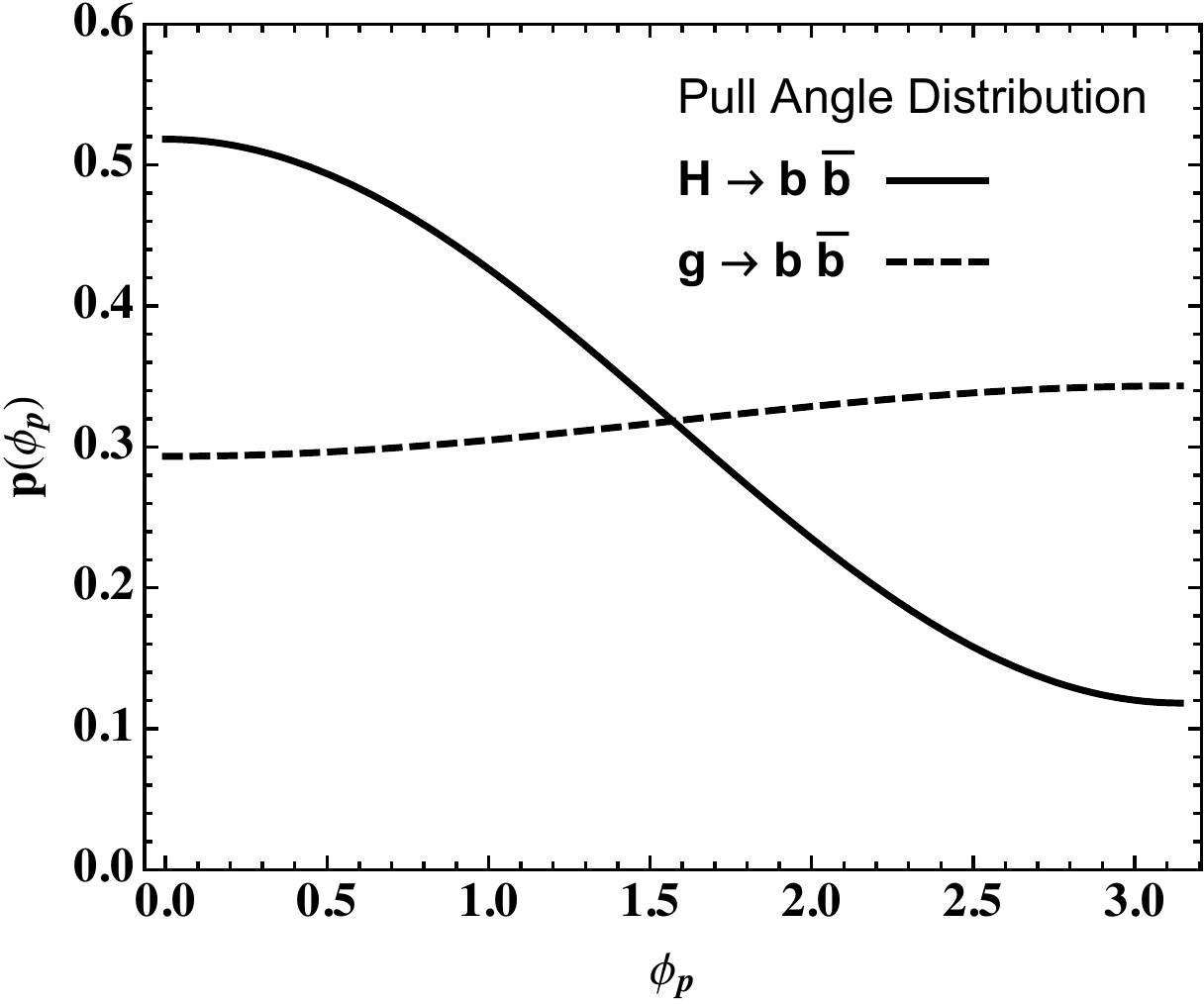}
\hspace{0.75cm}
\includegraphics[width=7cm]{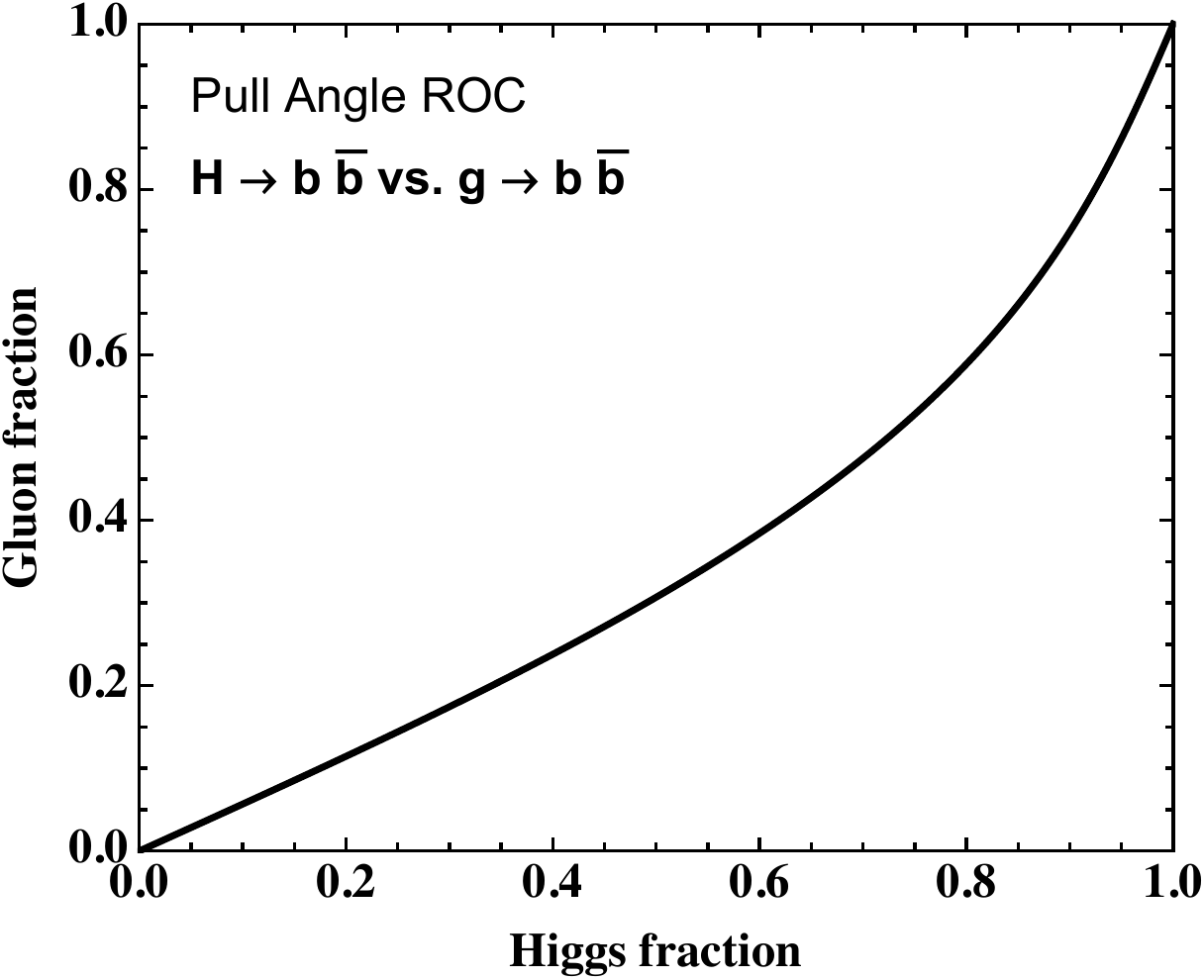}
\end{center}
\caption{(left) Distributions of the pull angle from Higgs decay and gluon splitting to bottom quarks in the high-boost and narrow jet limits.  We use $d_0 = 0.2$ to make the plots.  (right) Corresponding signal versus background efficiency curve for Higgs decay and gluon splitting to bottom quarks.
}
\label{fig:dists_roc}
\end{figure}

From these distributions, we can then quantify the discrimination power of the pull angle by making a sliding cut on the value of $\phi_p$.  The distributions of the pull angle from Higgs decay and gluon splitting are plotted in \Fig{fig:dists_roc}, where we use $d_0 = 0.2$.  On the right we plot the signal versus background efficiency curve or receiver operating characteristic (ROC) curve found from keeping those events that have pull angle below a sliding cut.  Because the pull angle peaks at small values for $H\to b\bar b$, this procedure amplifies the signal over the background, as manifest by the ROC curve lying below the diagonal.

To quantify the absolute power of the pull angle to discriminate Higgs from gluon splitting, we can integrate under the ROC curve.  This area-under-the-curve (AUC) vanishes for perfect discrimination and takes value $1/2$ for identical distributions.  The AUC can be calculated as an ordered integral over the two distributions and we find
\begin{align}
\text{AUC}_\text{(H vs.~g)} &= \int d\phi_s\int d\phi_b \, p_{H\to b\bar b}(\phi_s) \, p_{g\to b\bar b}(\phi_b)\, \Theta(\phi_s-\phi_b)=\frac{1}{2}-\frac{d_0}{\pi}\frac{C_A}{C_F} \\
&\simeq 0.35\nonumber\,.
\end{align}
The numerical value on the second line was found from setting $R = m_H/p_T$ and using the value of $d_0$ identified earlier.  For comparison, this value of the AUC is comparable to the value of the AUC for other jet discrimination problems, such as discriminating quark- from gluon-initiated jets.  For that problem, the discrimination power of the jet mass as quantified by the AUC at leading logarithmic accuracy is \cite{Larkoski:2013eya}
\begin{equation}
\text{AUC}_\text{($q$ vs.~$g$)} =  \frac{1}{1+\frac{C_A}{C_F}} \simeq 0.31\,.
\end{equation}
We also note, however, that these theoretical prediction of metrics may not be borne out in simulation or experiment, but are at least representative of the possible information available in the pull distribution for discrimination.

This simple calculation of course ignores many relevant physical effects that would exist in a real jet and would affect discrimination power.  The largest such effect would likely be from soft radiation uncorrelated or only weakly correlated with the direction of the jet.  Radiation that lands in the jet from color dipoles that are not color connected to the jet of interest would be approximately uniformly distributed over the area of the jet, with no preferred azimuthal direction.  The profile of such radiation in the pull angle $\phi_p$ would therefore be flat, and work to wash out the distinction between color singlet and octet dipoles studied in this section.  To mitigate this effect, one might groom the jet, removing soft, wide-angle radiation in the jet, but preserving radiation collinear to the jet axis.  However, collinear radiation is also flat in pull angle $\phi_p$ to lowest order in the collinear angular size, so this is also likely to wash out these subtle differences.  These considerations demonstrate the fragility of color correlations and possibly explain why pull has not been observed to be a useful discriminant in simulation \cite{ATLAS:2014rpa,deOliveira:2015xxd,Lin:2018cin}.

Beyond these practical considerations, one would like to have a formal understanding of the accuracy of the distributions and discrimination metrics derived in this section.  While the pull angle $\phi_p$ is not an IRC safe observable, its Sudakov safety means that one can vary renormalization scales in the integrand of \Eq{eq:ircunsafe} to have some estimate of theoretical uncertainties.  Such a procedure was also used in in \Ref{Tripathee:2017ybi} to estimate theoretical uncertainties for the groomed energy fraction $z_g$ \cite{Larkoski:2015lea}.  As observed in that case as well, we expect that this scale variation underestimates theoretical uncertainties on the calculated pull angle distribution.  Moving away from the small jet radius $R$ limit, we expect that increasing $R$ will likely improve discrimination power to a point.  More soft radiation that is sensitive to the dipole configuration will be included in the jets, but so too will more uniform contamination radiation, as mentioned earlier.  The effect of contamination radiation scales like the area of the jet, $R^2$, while the leading color-correlations between pairs of jets scales like $R$, so we expect there is some range of $R$ where contamination is small but color correlations are relatively large.  Further, higher-order effects like non-global logarithms \cite{Dasgupta:2001sh} will likely increase these color correlations present in the pull angle distribution.  Non-global effects will pull radiation in the jet toward ends of the dipoles that lie outside of the jet, in principle enhancing differences between the color singlet and octet configurations.

\section{Other Color Representations}\label{sec:colorreps}

The analysis of the previous section suggests a more general result for the pull distribution, appropriate for any combination of a pair of jets on which pull is measured.  We denote the jet 1 as the jet on which pull is measured and the jet 2 as the reference jet that defines the origin of the pull angle $\phi_p$.  In the limits in which the boost of the jet pair is large ($\theta_{12}\ll1$) and jet radius is small ($R\ll \theta_{12}$), the leading-order expression for the pull distribution is
\begin{align}
\frac{d^2\sigma^{R \ll \theta_{12}\ll 1}}{dt\, d\phi_p} &= \frac{\alpha_s}{\pi^2}\frac{1}{t}\left[{\bf T}_1^2\log\frac{R^2}{t}-{\bf T}_1^2B_1
+2\left[
{\bf T}_1^2+{\bf T}_2^2-({\bf T}_1+{\bf T}_2)^2
\right]\frac{R}{\theta_{12}}\cos\phi_p
\right]\,.
\end{align}
The coefficient of the term proportional to the jet radius $R$ is just another way to express the product of color matrices:
\begin{align}
{\bf T}_1^2+{\bf T}_2^2-({\bf T}_1+{\bf T}_2)^2 = -2{\bf T}_1\cdot{\bf T}_2\,.
\end{align}

We conjecture that this distribution is universal, in the limits $R\ll \theta_{12}\ll 1$ described above.  A unique aspect of this distribution is that there is a non-trivial term linear in the jet radius, $R$.  For many (if not nearly all) other observables, the first corrections to the distribution due to a finite jet radius are quadratic in $R$.  This is true of the jet mass, for example, and the magnitude of the pull vector, $t$.  The term proportional to $R$ integrates to 0 on $\phi_p\in[0,\pi]$, and so does not contribute to the pull magnitude's distribution.  The universality of this distribution along with the simplicity of color representations of pairs of jets in QCD enables us to explicitly enumerate all possible values for the quadratic Casimir difference, the coefficient of the $R$ term.  This is a concrete manifestation and justification for the name ``pull'': the difference of quadratic Casimirs explicitly corresponds to how soft radiation is pulled around the jets.  If the product representation is smaller than the sum of jets' Casimirs, then radiation is pulled between the pair of jets.  By contrast, if the product representation is larger than the pair of jets individually, radiation is pushed out of the pair.

A general irreducible representation of SU(3) can be represented with two non-negative integers $m_1$ and $m_2$ and denoted as $D(m_1,m_2)$.  The dimension for such a representation is
\begin{equation}
\text{dim}(m_1,m_2) =(1+m_1)(1+m_2)\left[
1+\frac{m_1+m_2}{2}
\right]\,.
\end{equation}
The quadratic Casimir of this representation is
\begin{equation}
C_{(m_1,m_2)} = {\bf T}^2_{(m_1,m_2)} =  \frac{1}{3}\left(m_1^2+m_2^2+3m_1+3m_2+m_1m_2\right)
\end{equation}
The fundamental and adjoint representations are ${\bf 3} = D(1,0)$ and ${\bf 8} = D(1,1)$, respectively, and these formulae give the correct values for the dimension and Casimir of these representations.  Exhaustive information about the representation theory of SU(3) can be found in \Ref{deSwart:1963pdg}.  The jets that form the pair on which pull is measured can only be some combination of quarks and gluons in QCD, so enumerating the possible product representations of color SU(3) that can appear is a simple exercise with SU(3) Clebsch-Gordan coefficients.

\begin{table}[t!]
\begin{center}
\begin{tabular}{ccc}
Jet Pair & Product Irrep & ${\bf T}_1^2+{\bf T}_2^2-({\bf T}_1+{\bf T}_2)^2$\\
\hline
$q\bar q$ & \begin{tabular}{@{}c@{}}{\bf 1}  \\ {\bf 8} \end{tabular} & \begin{tabular}{@{}c@{}}$\frac{8}{3}$  \\ $-\frac{1}{3}$ \end{tabular}\\
\hline
$qq$ & \begin{tabular}{@{}c@{}}$\bar{\bf 3}$  \\ ${\bf 6}=D(2,0)$ \end{tabular} & \begin{tabular}{@{}c@{}}$\frac{4}{3}$  \\ $-\frac{2}{3}$ \end{tabular}\\
\hline
$qg$ & \begin{tabular}{@{}c@{}}{\bf 3}  \\ $\bar{\bf 6}$\\${\bf 15}=D(2,1)$ \end{tabular} & \begin{tabular}{@{}c@{}} 3  \\ 1\\ $-1$ \end{tabular}\\
\hline
$gg$ & \begin{tabular}{@{}c@{}}{\bf 1}  \\ {\bf 8}\\${\bf 10}=D(3,0)$\\${\bf 27}=D(2,2)$ \end{tabular} & \begin{tabular}{@{}c@{}} 6  \\ 3 \\ 0\\ $-2$ \end{tabular}\\
\hline
\end{tabular}
\end{center}
\caption{Table of the possible jet pairs in QCD on which pull can be measured.  The irreps of color SU(3) that appear in the corresponding product representation are presented in the middle column.  We have only listed those irreps that correspond to unique values of the quadratic Casimir, e.g., {\bf 10} and $\overline{\bf 10}$ have the same quadratic Casimir.  In the final column, we calculate the ``pull'' of the product representation; the difference between the individual quadratic Casimirs of the two jets and their product representation.}\label{tab:pullreps}
\end{table}

In \Tab{tab:pullreps}, we list all possible QCD jet pairs, the irreps of SU(3) color that appear in their corresponding product representation, and then the value of the difference of quadratic Casimirs, using the formula presented earlier.  For most of the representations in the table, the difference of Casimirs is positive, indicating that the pull angle distribution peaks at $\phi_p=0$; that is, most radiation lies between the jet pair.  Only the highest dimension product representations produce negative Casimir differences, indicating that most radiation in this case is emitted outside of the region between the jet pair.  Intriguingly, the {\bf 10} representation of the color of a pair of gluons exhibits a perfectly flat pull angle distribution in this limit. Apparently this representation corresponds to exactly the same amount of radiation between as outside of the pair of gluon jets.

\section{Conclusions}\label{sec:concs}

The pull observable was designed to be sensitive to the flow of color between pairs of jets and thus sensitive to their product representation of SU(3) color.  This has been studied in simulation extensively and motivated measurements, but had not been justified theoretically.  In this paper, we demonstrated that pull, especially the pull angle, takes on a different distribution for pairs of jets in distinct product representations of color.  We performed explicit calculations at leading order for pull measured on $e^+e^-\to $ three jets events, studied the discrimination power of pull for identification of $H\to b\bar b$ decays, and presented a conjecture for the pull distribution in the high-boost, small jet radius limit.

The results presented in this paper suggest a number of extensions.  Observables that are more sensitive to color flow between jets can be designed, motivated by recent work in machine learning for particle physics \cite{Larkoski:2017jix,Guest:2018yhq,Radovic:2018dip}.  In particular, soft, wide-angle radiation in a jet is most sensitive to the colors of the other jets in an event, and to leading order, the distribution of this radiation is described by eikonal matrix elements.  With these eikonal matrix elements, one can construct the theoretically-optimal observable for discrimination of, say, a pair of jets that form a color singlet from a pair of jets that do not.  This optimal observable is the likelihood ratio by the Neyman-Pearson lemma \cite{Neyman:1933wgr}, and in general is not the pull angle.  Designing such observables may also resolve issues regarding residual color flow information in machine learning studies, even for jets on which pull is measured.

Prospects for observation of other hadronic decays of the Higgs boson could potentially be improved by using pull, or related color flow observables.  The Standard Model Higgs boson decays to pairs of gluons nearly 10\% of the time, and yet the $H\to gg$ decay mode is extremely challenging to observe.  Because gluons carry more color individually than quarks, the strength of color connection between the gluons in $H\to gg$ decays is substantially larger than between the bottom quarks in $H\to b\bar b$ decays, as shown in \Tab{tab:pullreps}.  This may suggest that it is easier to discriminate the $gg$ color singlet representation from non-singlet color representations; however, this may also mean that identification of the $H\to gg$ decay at high boost is more challenging to identify because the two hard prongs in the jet are less well-defined.

Finally, as the pull angle is not IRC safe, its calculation relies on resummed multi-differential cross sections to be well-defined.  Thus, ideally one would like the two-dimensional resummed cross section for the pull vector, from which the pull angle can be defined by marginalization.  How this resummation would proceed for different color configurations of jets would be interesting to determine.  Further, measurements of pull on $g\to b\bar b$ decay, for example, would test the extent to which the results derived in this paper were accurate at all at describing reality.  This then may point to a whole new class of observables that can be used to study global correlations in particle collision events.

\acknowledgments

A.~L.~thanks Simone Marzani and Chang Wu for collaboration on related projects and extensive discussions and Ben Nachman for comments on the draft.  Y.~B.~was supported by the physics research award from the Department of Physics at Reed College.

\appendix

\section{Non-connected Soft Function Calculation}\label{app:nonconsoft}

The double differential soft function for emissions from a dipole, neither of whose ends include the jet on which we measure the pull, can be calculated from
\begin{align}
S_{23}(t,\phi_p) = (-{\bf T}_2\cdot {\bf T}_3) g^2 \mu^{2\epsilon} \int [d^dk]_+ \frac{2 n_2\cdot n_3}{(k\cdot n_2)(k\cdot n_3)} \,\Theta_{\text{jet,}R}\, \delta_t\, \delta_{\phi_p}\,.
\end{align}
Here, we call the two ends of the dipole 2 and 3 and the phase space constraints $\Theta_{\text{jet,}R}$, $\delta_t$, and $\delta_{\phi_p}$ are the jet radius $R$ constraint, the measurement of the pull vector magnitude, and the measurement of the pull angle, respectively.  The light-like vectors $n_2$ and $n_3$ have unit 0th component and point along the direction of particles 2 and 3.  Three-jet production in $e^+e^-$ collisions in the center-of-mass frame is restricted to a plane, which simplifies the expression of the integrand.  With $\overline{\text{MS}}$ dimensional regularization, the soft function can be written as
\begin{align}
S_{23}(t,\phi_p)&=(-{\bf T}_2\cdot {\bf T}_3) \frac{\alpha_s}{\pi^{3/2}\Gamma(1/2-\epsilon)} \mu^{2\epsilon} \int_0^\infty dk_\perp \, k_\perp^{-1-2\epsilon}\int_{-\infty}^\infty d\eta \int_0^\pi d\phi \sin^{-2\epsilon}\phi\\
&
\hspace{1cm}
\times \frac{1-\cos\theta_{23}}{(\cosh\eta-\cos\phi\sin\theta_{12}-\sinh\eta\cos\theta_{12})(\cosh\eta+\cos\phi\sin\theta_{13}-\sinh\eta\cos\theta_{13})}\nonumber\\
&
\hspace{1cm}
\times \Theta\left(
\tan\frac{R}{2}-e^{-\eta}
\right)\,\delta(\phi_p-\phi)\, \delta\left(
t-\frac{k_\perp}{E_J \cosh\eta}
\right)\,.
\nonumber
\end{align}
Here, $E_J$ is the energy of the jet of interest, $\theta_{12}$ is the angle between the jet on which pull is measured and jet 2 (and correspondingly for $\theta_{13}$ and $\theta_{23}$).  Note the difference in sign in the $\cos\phi$ term in the two factors in the denominator of the matrix element term: jet 2 has an azimuthal angle of $\phi_2 = 0$ about jet 1 while jet 3 has an azimuthal angle of $\phi_3 = \pi$ about jet 1.  This assignment follows from the fact that jet 2 defines the location of the origin of the pull angle and that the collision occurs in the center-of-mass frame and the final state is confined to a plane.

The integrals that remain are finite for $\epsilon\to 0$, so we can just set $\epsilon = 0$ to calculate the corresponding pull distribution for $t>0$.  We find
\begin{align}
S_{23}(t,\phi_p)&=(-{\bf T}_2\cdot {\bf T}_3)\frac{\alpha_s}{\pi^2}\frac{1}{t}\frac{(\tan\frac{\theta_{12}}{2} + \tan\frac{\theta_{13}}{2})^2}{\tan^2\frac{\theta_{12}}{2}+\tan^2\frac{\theta_{13}}{2}+2\tan\frac{\theta_{12}}{2}\tan\frac{\theta_{13}}{2}\cos(2\phi_p)}  \\
&
\hspace{1cm}
\times\left[
\frac{\sin\frac{\theta_{12}-\theta_{13}}{2}}{\sin\frac{\theta_{12}+\theta_{13}}{2}}\log\left(
\frac{\frac{\tan^2\frac{R}{2}}{\tan^2\frac{\theta_{13}}{2}}+1+2\frac{\tan\frac{R}{2}}{\tan\frac{\theta_{13}}{2}}\cos\phi_p}{\frac{\tan^2\frac{R}{2}}{\tan^2\frac{\theta_{12}}{2}}+1-2\frac{\tan\frac{R}{2}}{\tan\frac{\theta_{12}}{2}}\cos\phi_p}
\right)\right.\nonumber\\
&\left.
\hspace{2cm}
+2\cot\phi_p \tan^{-1}\left(
\frac{\left(
\frac{\tan\frac{R}{2}}{\tan\frac{\theta_{12}}{2}}-\frac{\tan\frac{R}{2}}{\tan\frac{\theta_{13}}{2}}+2\frac{\tan\frac{R}{2}}{\tan\frac{\theta_{12}}{2}}\frac{\tan\frac{R}{2}}{\tan\frac{\theta_{13}}{2}} \cos\phi_p
\right)\sin\phi_p}{1-\left(
\frac{\tan\frac{R}{2}}{\tan\frac{\theta_{12}}{2}}-\frac{\tan\frac{R}{2}}{\tan\frac{\theta_{13}}{2}}
\right)\cos\phi_p-\frac{\tan\frac{R}{2}}{\tan\frac{\theta_{12}}{2}}\frac{\tan\frac{R}{2}}{\tan\frac{\theta_{13}}{2}}\cos(2\phi_p)}
\right)
\right]
\nonumber\\
&=(-{\bf T}_2\cdot {\bf T}_3)\frac{\alpha_s}{\pi^2}\frac{1}{t}\left(\frac{R^2 }{4}\left(\cot\frac{\theta_{12}}{2} + \cot\frac{\theta_{13}}{2}\right)^2+{\cal O}(R^3)\right)\,.\nonumber
\end{align}
In the final line, we have Taylor expanded the expression in powers of the jet radius $R$.  As expected, because there is no collinear singularity in this dipole configuration, the soft function is proportional to the area of the jet, $R^2$. One can also find a closed-form analytic expression for the more general case in which the jets 1, 2, and 3 do not lie in a plane, but we will not present it here.

\section{Gluon Jet Function}\label{app:gluejet}

To calculate the general distribution of pull measured on any jet in $e^+e^-\to$ three jets events, we must include the possibility of the jet being a gluon.  So, for the complete calculation, we also need to calculate the distribution of pull from collinear emissions in a gluon jet.  The $d=4-2\epsilon$, dimensionally-regulated, $\overline{\text{MS}}$ gluon jet function on which we measure pull is
\begin{align}
J_g(t,\phi_p)&=\frac{\alpha_s}{2\pi} \frac{1}{\pi^{1/2}\Gamma(1/2-\epsilon)}\left(
\frac{\mu^2}{E_J^2}
\right)^\epsilon\int_0^1dz\int_0^\infty d\theta^2\, (\theta^2)^{-1-\epsilon}\int_0^\pi d\phi\,\sin^{-2\epsilon}\phi\\
&
\hspace{0cm}
\times z^{-2\epsilon}(1-z)^{-2\epsilon}\left[
C_A\left(
\frac{1}{z}+\frac{1}{1-z}+z(1-z)-2
\right)+\frac{n_f}{2}\left(
1-\frac{2}{1-\epsilon}z(1-z)
\right)
\right]\nonumber\\
&
\hspace{0cm}
\times
\delta(t-z(1-z)|1-2z|\theta^2)\delta\left(
\phi_p-\phi
\right)\nonumber\,.
\end{align}
$n_f$ is the number of active fermions.  Performing the integrals over the $\delta$-functions, we then have
\begin{align}
J_g(t,\phi_p)&=\frac{\alpha_s}{2\pi} \frac{1}{\pi^{1/2}\Gamma(1/2-\epsilon)}\left(
\frac{\mu^2}{E_J^2\sin^2\phi_p}
\right)^\epsilon\frac{1}{t^{1+\epsilon}}\int_0^1dz\, z^{-1-\epsilon}\\
&
\hspace{0cm}
\times (1-z)^{-\epsilon}|1-2z|^\epsilon\left[
C_A\left(
2+z^2(1-z)-2z
\right)+\frac{n_f}{2}\left(
z-\frac{2}{1-\epsilon}z^2(1-z)
\right)
\right]\nonumber\,.
\end{align}
In this expression, we have also symmetrized the first two terms of the splitting function, to isolate the divergence at $z=0$.  To integrate over $z$, we can expand the first factor in $+$-functions:
\begin{equation}
z^{-1-\epsilon} = -\frac{1}{\epsilon}\delta(z) + \left(\frac{1}{z}\right)_+ +\cdots\,.
\end{equation}
The integral with the $\delta(z)$ is just $2 C_A$.  For the integral with the $+$-function, we can set $\epsilon = 0$ and we have
\begin{align}
&\int_0^1dz\, \left(\frac{1}{z}\right)_+\left[
C_A\left(
2+z^2(1-z)-2z
\right)+\frac{n_f}{2}\left(
z-2z^2(1-z)
\right)
\right]\\
&
\hspace{2cm}
=\int_0^1dz\, \left[
C_A\left(
z(1-z)-2
\right)+\frac{n_f}{2}\left(
1-2z(1-z)
\right)
\right]\nonumber\\
&\hspace{2cm}
=-\frac{11}{6}C_A+\frac{n_f}{3}\,.\nonumber
\end{align}

With these results, the jet function is
\begin{align}
J_g(t,\phi_p)&=\frac{\alpha_s C_A}{\pi} \frac{1}{\pi^{1/2}\Gamma(1/2-\epsilon)}\left(
\frac{\mu^2}{E_J^2\sin^2\phi_p}
\right)^\epsilon\frac{1}{t^{1+\epsilon}}\left[-\frac{1}{\epsilon}-\frac{11}{12}+\frac{n_f}{6 C_A}
\right]\,.
\end{align}
Only keeping those terms that contribute for $t>0$, the jet function is then
\begin{equation}
J_g(t,\phi_p) = \frac{\alpha_s C_A}{\pi^2}\frac{1}{t}\left[
\log\frac{4t E_J^2 \sin^2\phi_p}{\mu^2}-\frac{11}{12}+\frac{n_f}{6 C_A}
\right]\,.
\end{equation}

\bibliography{pull}

\providecommand{\href}[2]{#2}\begingroup\raggedright\begin{thebibliography}{10}

\bibitem{Field:1977fa}
R.~D. Field and R.~P. Feynman, {\it {A Parametrization of the Properties of
  Quark Jets}},  {\em Nucl. Phys.} {\bf B136} (1978) 1 -- 76.

\bibitem{Krohn:2012fg}
D.~Krohn, M.~D. Schwartz, T.~Lin, and W.~J. Waalewijn, {\it {Jet Charge at the
  {LHC}}},  {\em Phys. Rev. Lett.} {\bf 110} (2013), no.~21 212001,
  [\href{http://arxiv.org/abs/1209.2421}{{\tt arXiv:1209.2421}}].

\bibitem{Hanson:1975fe}
G.~Hanson et~al., {\it {Evidence for Jet Structure in Hadron Production by e+
  e- Annihilation}},  {\em Phys. Rev. Lett.} {\bf 35} (1975) 1609 -- 1612.

\bibitem{Brandelik:1979bd}
{\bf TASSO} Collaboration, R.~Brandelik et~al., {\it {Evidence for Planar
  Events in e+ e- Annihilation at High-Energies}},  {\em Phys. Lett.} {\bf B86}
  (1979) 243 -- 249.

\bibitem{Barber:1979yr}
D.~P. Barber et~al., {\it {Discovery of Three Jet Events and a Test of Quantum
  Chromodynamics at {PETRA}}},  {\em Phys. Rev. Lett.} {\bf 43} (1979) 830 --
  833.

\bibitem{Berger:1979cj}
{\bf PLUTO} Collaboration, C.~Berger et~al., {\it {Evidence for Gluon
  Bremsstrahlung in e+ e- Annihilations at High-Energies}},  {\em Phys. Lett.}
  {\bf B86} (1979) 418 -- 425.

\bibitem{Bartel:1979ut}
{\bf JADE} Collaboration, W.~Bartel et~al., {\it {Observation of Planar Three
  Jet Events in e+ e- Annihilation and Evidence for Gluon Bremsstrahlung}},
  {\em Phys. Lett.} {\bf 91B} (1980) 142 -- 147.

\bibitem{Gallicchio:2010sw}
J.~Gallicchio and M.~D. Schwartz, {\it {Seeing in Color: Jet Superstructure}},
  {\em Phys. Rev. Lett.} {\bf 105} (2010) 022001,
  [\href{http://arxiv.org/abs/1001.5027}{{\tt arXiv:1001.5027}}].

\bibitem{Abazov:2011vh}
{\bf D0} Collaboration, V.~M. Abazov et~al., {\it {Measurement of Color Flow in
  $\mathbf{t\bar{t}}$ Events from $\mathbf{p\bar{p}}$ Collisions at
  $\mathbf{\sqrt{s}=1.96}$ TeV}},  {\em Phys. Rev.} {\bf D83} (2011) 092002,
  [\href{http://arxiv.org/abs/1101.0648}{{\tt arXiv:1101.0648}}].

\bibitem{Aad:2015lxa}
{\bf ATLAS} Collaboration, G.~Aad et~al., {\it {Measurement of colour flow with
  the jet pull angle in $t\bar{t}$ events using the {ATLAS} detector at
  $\sqrt{s}=8$ TeV}},  {\em Phys. Lett.} {\bf B750} (2015) 475 -- 493,
  [\href{http://arxiv.org/abs/1506.05629}{{\tt arXiv:1506.05629}}].

\bibitem{Aaboud:2018ibj}
{\bf ATLAS} Collaboration, M.~Aaboud et~al., {\it {Measurement of colour flow
  using jet-pull observables in $t\bar{t}$ events with the ATLAS experiment at
  $\sqrt{s} = 13\,\hbox {TeV}$}},  {\em Eur. Phys. J.} {\bf C78} (2018), no.~10
  847, [\href{http://arxiv.org/abs/1805.02935}{{\tt arXiv:1805.02935}}].

\bibitem{Chatrchyan:2012xdj}
{\bf CMS} Collaboration, S.~Chatrchyan et~al., {\it {Observation of a New Boson
  at a Mass of 125 GeV with the {CMS} Experiment at the {LHC}}},  {\em Phys.
  Lett.} {\bf B716} (2012) 30 -- 61,
  [\href{http://arxiv.org/abs/1207.7235}{{\tt arXiv:1207.7235}}].

\bibitem{Chatrchyan:2013zna}
{\bf CMS} Collaboration, S.~Chatrchyan et~al., {\it {Search for the standard
  model Higgs boson produced in association with a W or a Z boson and decaying
  to bottom quarks}},  {\em Phys. Rev.} {\bf D89} (2014), no.~1 012003,
  [\href{http://arxiv.org/abs/1310.3687}{{\tt arXiv:1310.3687}}].

\bibitem{Chatrchyan:2014tja}
{\bf CMS} Collaboration, S.~Chatrchyan et~al., {\it {Search for invisible
  decays of Higgs bosons in the vector boson fusion and associated ZH
  production modes}},  {\em Eur. Phys. J.} {\bf C74} (2014) 2980,
  [\href{http://arxiv.org/abs/1404.1344}{{\tt arXiv:1404.1344}}].

\bibitem{Khachatryan:2014dea}
{\bf CMS} Collaboration, V.~Khachatryan et~al., {\it {Measurement of
  electroweak production of two jets in association with a Z boson in
  proton-proton collisions at $\sqrt{s}=8\,\text{TeV}$}},  {\em Eur. Phys. J.}
  {\bf C75} (2015), no.~2 66, [\href{http://arxiv.org/abs/1410.3153}{{\tt
  arXiv:1410.3153}}].

\bibitem{Khachatryan:2016mdm}
{\bf CMS} Collaboration, V.~Khachatryan et~al., {\it {Search for dark matter in
  proton-proton collisions at 8 TeV with missing transverse momentum and vector
  boson tagged jets}},  {\em JHEP} {\bf 12} (2016) 083,
  [\href{http://arxiv.org/abs/1607.05764}{{\tt arXiv:1607.05764}}]. [Erratum:
  JHEP08,035(2017)].

\bibitem{Larkoski:2019urm}
A.~J. Larkoski, S.~Marzani, and C.~Wu, {\it {Theory Predictions for the Pull
  Angle}},  {\em Phys. Rev.} {\bf D99} (2019), no.~9 091502,
  [\href{http://arxiv.org/abs/1903.02275}{{\tt arXiv:1903.02275}}].

\bibitem{ATLAS:2014rpa}
{\bf ATLAS} Collaboration, T.~A. collaboration, {\it {Reconstruction and
  Modelling of Jet Pull with the {ATLAS} Detector}}, .

\bibitem{deOliveira:2015xxd}
L.~de~Oliveira, M.~Kagan, L.~Mackey, B.~Nachman, and A.~Schwartzman, {\it
  {Jet-images -- deep learning edition}},  {\em JHEP} {\bf 07} (2016) 069,
  [\href{http://arxiv.org/abs/1511.05190}{{\tt arXiv:1511.05190}}].

\bibitem{Lin:2018cin}
J.~Lin, M.~Freytsis, I.~Moult, and B.~Nachman, {\it {Boosting $H\to b\bar b$
  with Machine Learning}},  {\em JHEP} {\bf 10} (2018) 101,
  [\href{http://arxiv.org/abs/1807.10768}{{\tt arXiv:1807.10768}}].

\bibitem{Aad:2019uoz}
{\bf ATLAS} Collaboration, G.~Aad et~al., {\it {Identification of boosted Higgs
  bosons decaying into $b$-quark pairs with the ATLAS detector at 13 TeV}},
  {\em Eur. Phys. J.} {\bf C79} (2019), no.~836
  [\href{http://arxiv.org/abs/1906.11005}{{\tt arXiv:1906.11005}}].

\bibitem{Larkoski:2013paa}
A.~J. Larkoski and J.~Thaler, {\it {Unsafe but Calculable: Ratios of
  Angularities in Perturbative {QCD}}},  {\em JHEP} {\bf 09} (2013) 137,
  [\href{http://arxiv.org/abs/1307.1699}{{\tt arXiv:1307.1699}}].

\bibitem{Larkoski:2015lea}
A.~J. Larkoski, S.~Marzani, and J.~Thaler, {\it {Sudakov Safety in Perturbative
  {QCD}}},  {\em Phys. Rev.} {\bf D91} (2015), no.~11 111501,
  [\href{http://arxiv.org/abs/1502.01719}{{\tt arXiv:1502.01719}}].

\bibitem{Larkoski:2013eya}
A.~J. Larkoski, G.~P. Salam, and J.~Thaler, {\it {Energy Correlation Functions
  for Jet Substructure}},  {\em JHEP} {\bf 06} (2013) 108,
  [\href{http://arxiv.org/abs/1305.0007}{{\tt arXiv:1305.0007}}].

\bibitem{Tripathee:2017ybi}
A.~Tripathee, W.~Xue, A.~Larkoski, S.~Marzani, and J.~Thaler, {\it {Jet
  Substructure Studies with {CMS} Open Data}},  {\em Phys. Rev.} {\bf D96}
  (2017), no.~7 074003, [\href{http://arxiv.org/abs/1704.05842}{{\tt
  arXiv:1704.05842}}].

\bibitem{Dasgupta:2001sh}
M.~Dasgupta and G.~P. Salam, {\it {Resummation of nonglobal {QCD}
  observables}},  {\em Phys. Lett.} {\bf B512} (2001) 323 -- 330,
  [\href{http://arxiv.org/abs/hep-ph/0104277}{{\tt hep-ph/0104277}}].

\bibitem{deSwart:1963pdg}
J.~J. de~Swart, {\it {The Octet model and its Clebsch-Gordan coefficients}},
  {\em Rev. Mod. Phys.} {\bf 35} (1963) 916 -- 939. [Erratum: Rev. Mod.
  Phys.37,326(1965)].

\bibitem{Larkoski:2017jix}
A.~J. Larkoski, I.~Moult, and B.~Nachman, {\it {Jet Substructure at the Large
  Hadron Collider: A Review of Recent Advances in Theory and Machine
  Learning}},  \href{http://arxiv.org/abs/1709.04464}{{\tt arXiv:1709.04464}}.

\bibitem{Guest:2018yhq}
D.~Guest, K.~Cranmer, and D.~Whiteson, {\it {Deep Learning and its Application
  to {LHC} Physics}},  {\em Ann. Rev. Nucl. Part. Sci.} {\bf 68} (2018) 161 --
  181, [\href{http://arxiv.org/abs/1806.11484}{{\tt arXiv:1806.11484}}].

\bibitem{Radovic:2018dip}
A.~Radovic, M.~Williams, D.~Rousseau, M.~Kagan, D.~Bonacorsi, A.~Himmel,
  A.~Aurisano, K.~Terao, and T.~Wongjirad, {\it {Machine learning at the energy
  and intensity frontiers of particle physics}},  {\em Nature} {\bf 560}
  (2018), no.~7716 41 -- 48.

\bibitem{Neyman:1933wgr}
J.~Neyman and E.~S. Pearson, {\it {On the Problem of the Most Efficient Tests
  of Statistical Hypotheses}},  {\em Phil. Trans. Roy. Soc. Lond.} {\bf A231}
  (1933), no.~694-706 289 -- 337.

\end{thebibliography}\endgroup

\end{document}